\documentclass[11pt]{article}
\usepackage[a4paper,hmargin=1.0in,vmargin=1.0in]{geometry}
\usepackage{amsmath,amsthm,amssymb,sectsty,setspace,color,comment,graphicx}   
\usepackage{optidef}
\usepackage[numbers,square]{natbib}

\usepackage{xcolor}
\definecolor{light-gray}{gray}{0.95}

\usepackage[hidelinks]{hyperref}   % makes URLs clickable
\usepackage{url}        % improves URL handling
\usepackage{xurl}       % (optional) allows line breaks at ANY character if needed

\usepackage{float}

\theoremstyle{plain}

\newtheorem{definition}{Definition}[section]

\sectionfont{\large} \subsectionfont{\normalsize}
\allowdisplaybreaks
\begin{document}

\begin{titlepage}

\title{The Data Enclave Advantage: A New Paradigm for Least-Privileged Data Access in a Zero-Trust World}
%\subtitle{A New Paradigm for Least-Privileged Data Access in a Zero-Trust World}
\author{
Nico Bistolfi\thanks{AlienGiraffe AI,
\{nico,andreea\}@aliengiraffe.ai.}%
\and
Andreea Georgescu\footnotemark[1]%
\and
Dave Hodson\thanks{Advisor at AlienGiraffe AI, dave@davehodson.com}%
}
\date{}
\maketitle

\thispagestyle{empty}

\begin{abstract}
As cloud infrastructure evolves to support dynamic and distributed workflows, accelerated now by AI-driven processes, the outdated model of standing permissions has become a critical vulnerability. Based on the Cloud Security Alliance (CSA) Top Threats to Cloud Computing Deep Dive 2025 Report \cite{csa_2025_report}, our analysis details how standing permissions cause catastrophic cloud breaches. While current security tools are addressing network and API security, the challenge of securing granular data access remains. Removing standing permissions at the data level is as critical as it is at the network level, especially for companies handling valuable data at scale.

In this white paper, we introduce an innovative architecture based on \textbf{on-demand data enclaves} to address this gap directly. Our approach enables Zero Standing Privilege (ZSP) and Just-in-Time (JIT) principles at the data level. We replace static permissions with temporary data contracts that enforce proactive protection. This means separation is built around the data requested on-demand, providing precise access and real time monitoring for individual records instead of datasets. This solution drastically reduces the attack surface, prevents privilege creep, and simplifies auditing, offering a vital path for enterprises to transition to a more secure and resilient data environment.

\end{abstract}

\medskip \noindent {\bf Keywords:}  Database, Zero Trust, Standing Permissions, Just in Time Permissions, Data Contract, Containers, Data Leak, Cloud Computing

\end{titlepage}

% CONTENTS %%%%%%%%%%%%%%%%%%%%%%%%%%%%%%%%%%%%%%
\tableofcontents
\thispagestyle{empty}
\newpage

% INTRO %%%%%%%%%%%%%%%%%%%%%%%%%%%%%%%%%%%%%%%%%
\setcounter{page}{1}

\section{Introduction}

\subsection{The Problem with Standing Permissions}

The security of cloud infrastructure has become the single most critical concern for organizations of all sizes. While advancements in cloud security tools and practices have made significant strides, a fundamental vulnerability continues to be a leading cause of major data breaches and operational failures: the reliance on standing permissions. Standing permissions are broad scoped and long-lived access rights that remain active for an indefinite period, often long after the initial task or role for which they were granted has been completed. This model, a holdover from less-dynamic, on-premises environments, is catastrophically unsuited for the fluid and complex nature of the cloud.

The Cloud Security Alliance Top Threats to Cloud Computing Deep Dive 2025 Report \cite{csa_2025_report} provides a crucial, evidence-based look at this problem. By analyzing high-profile security incidents, the report identifies a clear and alarming trend: Identity and Access Management (IAM) failures, often rooted in broad scoped standing permissions, are the most frequent threat observed in cloud breaches. 

We expand on the report's findings, exploring why standing permissions are so dangerous and why a transition to a Zero Standing Privilege (ZSP) and Just-in-Time (JIT) access model is no longer just a best practice, but a business imperative for survival in the cloud. We will dissect the mechanisms through which standing permissions enable attacks, the challenges they pose for governance, and the strategic solutions that can help organizations build a more secure future.

\subsection{The Challenge of Applying Least Privilege to Data}

While tools like Privileged Access Management (PAM) and Identity and Access Management (IAM) are helping companies move toward a ZSP and JIT model for network and API security, data security is much more intricate. Current methods, such as Role-Based Access Control (RBAC), Row-Level Security (RLS), programmatic access controls and database proxies, are based on "standing permissions" that don't allow for true "least privilege" access. 

These static controls are often too broad in order to avoid slowing down data access, and they lack continuous monitoring and JIT architecture. The core challenge is shifting from controlling broad database connections to dynamically managing access to individual data records precisely when they're needed for a specific query.

\subsection{Introducing the Data Enclave}

We propose an architecture to enable  ZSP and JIT practices to the data level, with minimal architecture changes. The latter being particularly important for legacy enterprises where a full architecture is impossible. But also for the age of AI, where flows are becoming more dynamic and unpredictable, so the goal of setting up separation a priori is not practical.

Instead, our architecture revolves around on-demand data enclaves. A Data Enclave is a secure, isolated environment created on-demand to provide just-in-time access to a specific subset of data. This model replaces "standing permissions" with temporary "data contracts" that define precisely who can access what data, for how long, and for what purpose. As well as monitoring queries in real time. By doing this, enclaves significantly reduce the attack surface, prevent privilege creep, and simplify the auditing process.

\section{Standing permissions in recent incidents}

\subsection{Expanded Attack Surface}
Standing permissions directly contribute to an expanded attack surface, turning a single compromised identity into a master key for an entire environment. Instead of a narrow point of entry, attackers are presented with a landscape of opportunities. This is clearly demonstrated in the Darkbeam incident \cite{darkbeam_incident}, where a misconfigured Elasticsearch and Kibana interface was left publicly accessible without any authentication. While this was a misconfiguration, the underlying issue was a lack of a ZSP model. The system had, in effect, standing permission for anyone on the internet to access it, exposing over \$3.8 billion worth of email-password combinations.

This expanded attack surface is the primary reason why breaches can go undetected for extended periods. As seen with the Toyota breach, a simple misconfiguration that granted standing public access to sensitive data persisted for nearly a decade. The absence of a least-privilege policy for the cloud environment meant that any user or service with access had the potential to create a massive data leak. The problem wasn't a malicious actor, but the standing permissions that created a vulnerability and a wide-open attack surface. The longer these permissions exist, the more likely they are to be discovered and exploited by an external or internal actor.

\subsection{Privilege Creep}
Standing permissions are the root cause of "privilege creep," a phenomenon where users and applications accumulate more permissions over time than they actually need to perform their jobs. This occurs because it is easier to grant new permissions than to revoke old ones, especially as roles and responsibilities change within an organization. Over time, an administrator might gain permissions to a legacy system, a new database, and a third-party application, and these permissions remain active long after they are necessary.

The Microsoft and Retool \& Fortress breaches \cite{microsoft_incident}, \cite{retool_incident} provide textbook examples of how privilege creep can be exploited. In the Microsoft case, the attacker's initial access to a "legacy, non-production test tenant account" was catastrophic precisely because that account had amassed elevated privileges over time, allowing the attacker to create new malicious applications and escalate their access. Similarly, the Retool \& Fortress breach, initiated through a social engineering attack on an employee, was made possible because the compromised account had excessive standing privileges that enabled the attacker to access multiple applications and modify sensitive customer data. In both scenarios, an attacker leveraged existing, unnecessary standing permissions to move laterally and escalate their access, demonstrating how privilege creep turns a minor security incident into a full-scale breach.

\subsection{Lateral Movement and Privilege Escalation}
After gaining an initial foothold, a threat actor's next objective is to move deeper into the compromised environment—a process known as lateral movement—and gain higher levels of access—privilege escalation. Standing permissions are the key enablers of this phase of an attack. When a compromised account already has broad, long-lived access to multiple systems, an attacker doesn't need to perform a series of complex exploits. They can simply use the existing permissions to navigate the network, escalate privileges, and find their target data.

The Microsoft breach \cite{microsoft_incident} is a prime example of this. The attacker's initial access to a low-privilege test account was not a dead end. Instead, that account had standing access to an OAuth application with elevated permissions, which in turn allowed the attacker to create additional malicious OAuth applications and access corporate mailboxes. The attacker used the pre-existing permissions to move from an isolated test environment into the core of Microsoft's corporate network.

Similarly, the FTX collapse \cite{ftx_incident}, fueled by a SIM-swap attack, was a lesson in how a lack of internal segmentation combined with standing permissions can facilitate rapid lateral movement. Once the attackers compromised an account, they were able to exploit "poor key management and insufficient internal segmentation, allowing lateral movement and broad access to funds." The standing permissions on the exchange’s hot wallets meant that once the attacker was in, they had the keys to the kingdom. Had the funds been held in a system with limited-time access, or had the keys been protected by a JIT model, the attacker's ability to move and exfiltrate funds would have been severely hindered.

The Snowflake breach \cite{snowflake_incident} further demonstrates how standing permissions enable lateral movement. The attackers used credentials stolen via infostealer malware to gain access to customer accounts. The absence of MFA and other security controls meant that these standing credentials were all the attackers needed. Once authenticated, they could perform a series of queries and commands \colorbox{light-gray}{\texttt{SHOW TABLES, SELECT * FROM, COPY INTO}} to stage and exfiltrate terabytes of sensitive data. They didn't need to perform any further privilege escalation; the standing permissions were all they needed to achieve their objective.

\subsection{Auditing Challenges}
The persistence of standing permissions creates immense challenges for security auditing and compliance. The sheer volume of permissions granted in a typical cloud environment can make it nearly impossible for human auditors to track, verify, and validate that every permission is justified and actively used. The "Top Threats" report emphasizes the need for continuous monitoring and a "Detection of Baseline Deviation" (CCC-07). However, when standing permissions are the baseline, this control becomes less effective.

The Toyota data leak \cite{toyota_incident}, which persisted for ten years, and the Darkbeam incident, which was discovered by an external researcher, both highlight a fundamental failure in auditing. The lack of routine audits and strategic oversight of cloud configurations allowed these dangerous standing permissions to persist, and the organizations failed to detect the public exposure themselves. The report's analysis of both incidents points to insufficient logging and monitoring as key vulnerabilities. Without a dynamic model, organizations are forced to rely on infrequent, manual audits that are easily overwhelmed by the complexity of cloud environments, leaving them blind to critical security gaps.

\subsection{The “Business Justification” for Breaking Glass}
A common argument for maintaining standing permissions is the need for "break-glass" access in an emergency. The thinking is that in a critical situation—like a service outage or a security incident—a small group of highly privileged accounts must have immediate, standing access to perform emergency remediation. However, this traditional approach is fundamentally flawed and, as the report shows, often contributes to the very incidents it is meant to solve. The Microsoft breach, for example, was a result of an attacker leveraging an account with exactly this type of elevated access.

A more secure and effective "break-glass" strategy, in line with the principles of least privilege and ZSP, involves a carefully orchestrated, JIT-based process. This involves:

\begin{itemize}
  \item \textbf{Strictly defined and secured "break-glass" accounts}: These accounts have no standing permissions, used exclusively to request temporary, highly-privileged access.
  \item \textbf{Multi-person approval}: The request for "break-glass" access requires approval from multiple authorized individuals.
  \item \textbf{Automated activation and deactivation}: The permissions are automatically activated for a very short, pre-defined period and then automatically revoked.
  \item \textbf{Real-time monitoring and alerting}: The activation of a "break-glass" account triggers immediate, high-priority alerts to the security team, ensuring that all actions taken are fully logged, monitored, and auditable in real-time.
\end{itemize}

This approach ensures that while the business can still respond to emergencies, it does so in a way that minimizes risk rather than creating it. The FTX and Microsoft cases \cite{microsoft_incident}, \cite{ftx_incident} both underscore the importance of cloud-specific incident response plans (SEF-03). A modern, robust plan must include a secure "break-glass" protocol that eliminates standing permissions.

\section{Data vs. Network and other permissions}

The concepts presented are widely recognized within the cybersecurity domain. While there is broad consensus regarding the "ideal" state, the practical implementation poses significant challenges. The feasibility of achieving this ideal is often assessed by the associated costs and the degree of friction introduced.

In the realm of network security and API-defined permissions (e.g., read vs. write operations) and thinking about an implementation of SDP, tools such as Privileged Access Management (PAM) and Identity and Access Management (IAM) are gradually guiding organizations toward this optimal state - enabling Zero Trust solutions to the resource level.

However, data presents a more intricate challenge. Current organizational practices involve perimeter-based data protection, database segregation, Role Based Access (RBAC) at the table and row level (RLS), programmatic access controls, and the utilization of proxies. These methods, in essence, constitute "standing permissions”. RBAC and RLS provide more granular controls, but the static and manual nature prohibits true “least privileged” access. Organizations often create broad roles and row/column controls for fear of adding friction and slowing down data access. Similarly, programmatic access control lacks automations and the infrastructure to ensure it’s always enforced. And all these methods lack continuous monitoring and JIT access. 

As illustrated by the Snowflake example \cite{snowflake_incident}, compromising credentials for an appropriate role, accessing a table, and executing a SQL injection beyond the proxy can lead to a \colorbox{light-gray}{\texttt{SELECT *}} operation. In practice, comprehensive data access is often unnecessary. Rather, access is typically limited to the most recent timeframe, a specific account, or a subset of columns.

The inherent difficulty lies in controlling individual records rather than merely database connections, and in dynamically applying "separation" just-in-time, around the data required for query execution.

\newpage
\section{A Solution: Zero-Trust Data Enclaves}

As organizations transition away from standing permissions, a strategic shift to a Zero-Trust architecture becomes imperative. A key component of this architecture is the use of data enclaves to provision access. This approach ensures that data access is never granted in a blanket fashion but is instead brokered through a secure, isolated environment - emulating the works of a man trap, one door needs to be closed for the other to open.

\begin{definition}
A Zero-Trust Data Enclave is a secure, isolated environment with its own set of finely tuned controls. It houses a subset of an organization's sensitive data, along with the specific compute, storage, and networking resources needed to process it. Access to this enclave is strictly governed by a pre-authorized agreement that specifies exactly which users or applications can access what data, for how long, and for what purpose. We’ll call these, “data contracts”.
\end{definition}

\subsection{Data Enclaves work like a "man trap" for data}

\textbf{Step 1} - At this point, the Data Enclave already knows which data contract to implement, what resources to connect to and which segments of data to pull from each resource.

\begin{figure}[H]
    \centering
    \includegraphics[scale = 0.65]{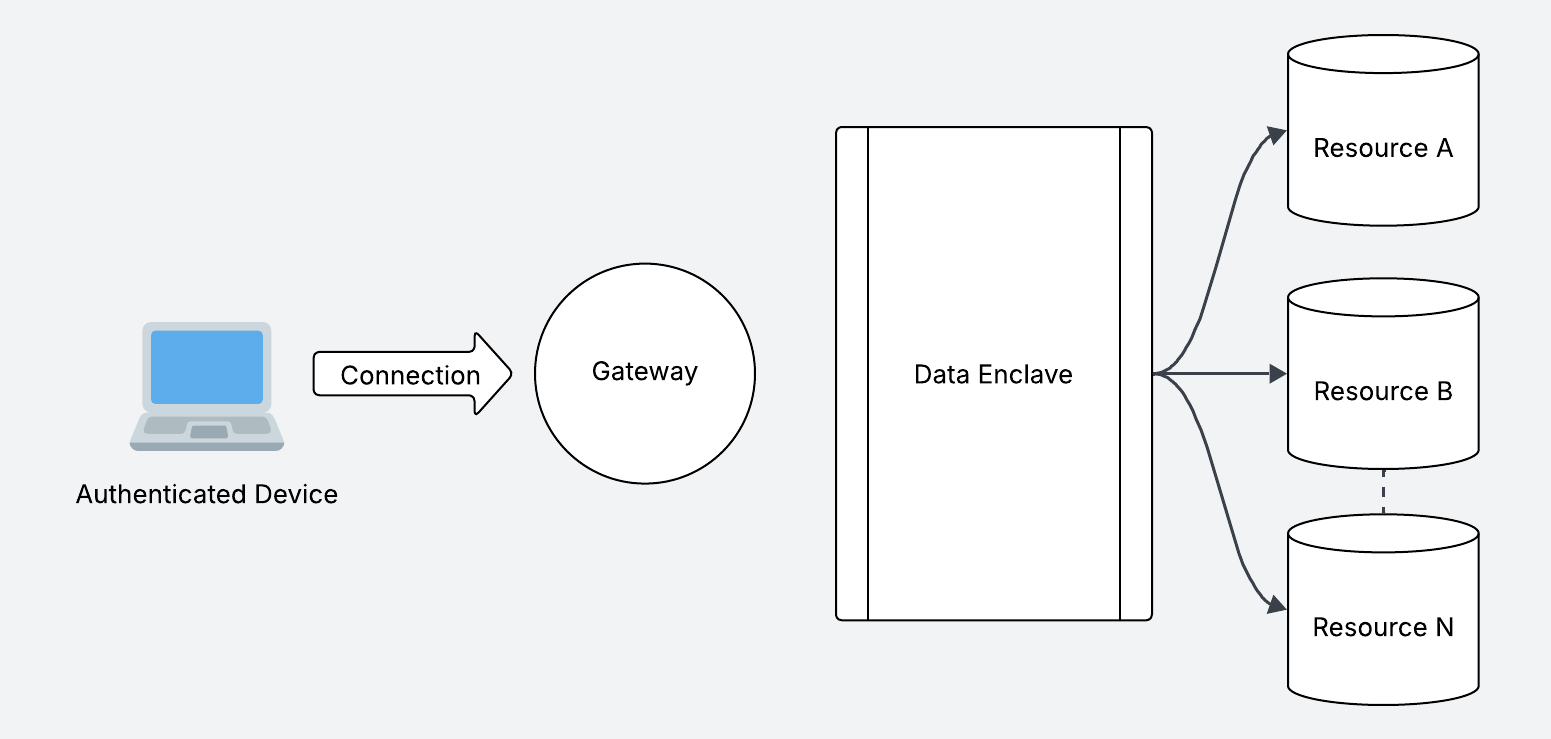}
    \caption{The Data Enclave implements contracts at the resource and data segment level.}
    \label{fig:implement_contract}
\end{figure}

\textbf{Step 2} - The Data Enclave disconnects from the resources, and allows the Gateway managing the connection with the host, to connect and query the data.

\begin{figure}[H]
    \centering
    \includegraphics[scale = 0.65]{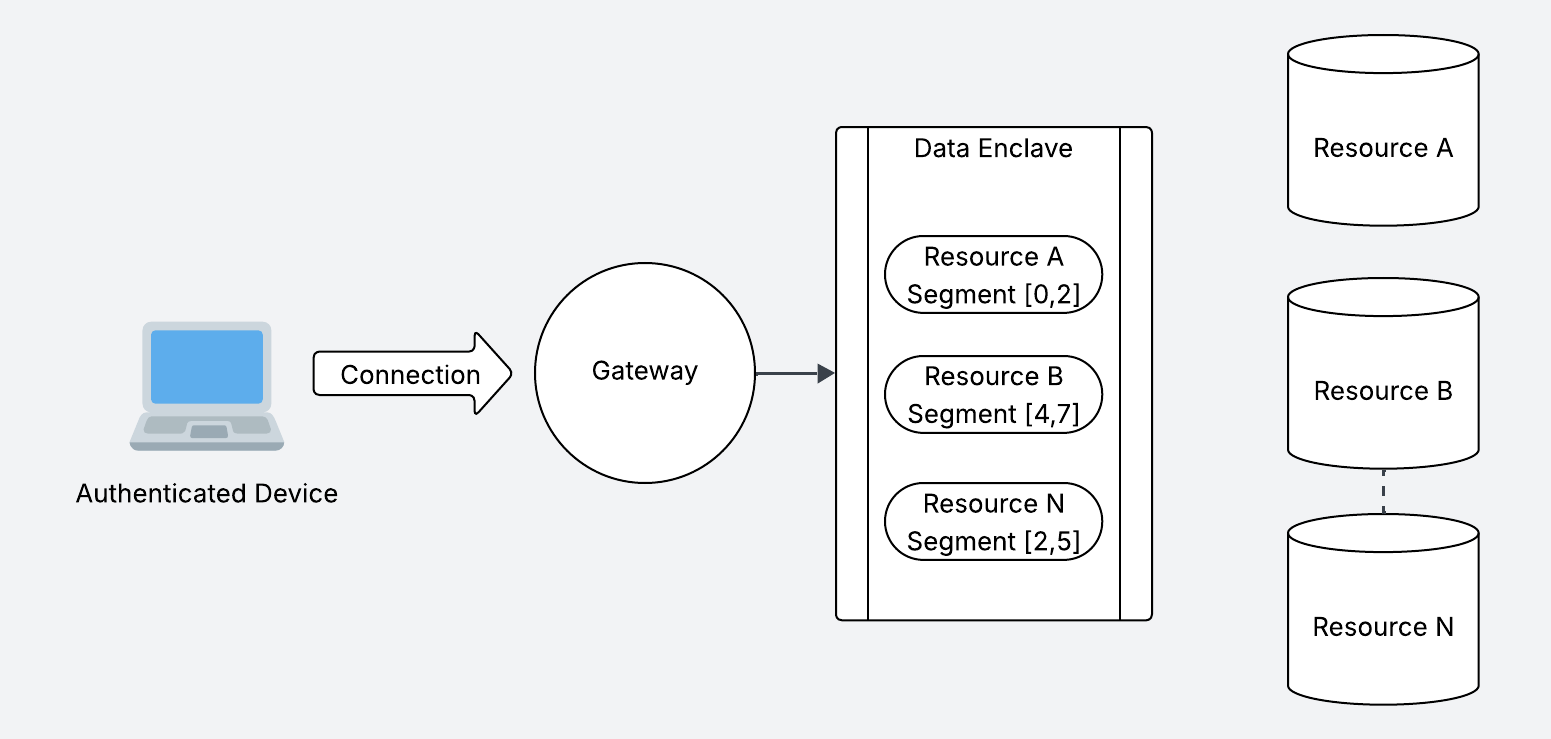}
    \caption{The Data Enclave disconnects from the resources, to achieve physical separation. Keep in mind that this is a simplified and conceptual version, a complete implementation diagram would have more elements and the complexity is out of scope for this white paper.}
    \label{fig:choice_models}
\end{figure}

This model solves the problems of standing permissions in several key ways:

\begin{itemize}
    \item \textbf{Eliminates Expanded Attack Surface}: By their very nature, enclaves limit the blast radius of a breach. An attacker who compromises a user or application is confined to that specific enclave and the data contract it permits. Lateral movement to other parts of the network—such as a different enclave with a different data set or a different user's data—is impossible without a new, explicit data contract.
    \item \textbf{Combats Privilege Creep}: Unlike a traditional access model where a single user's permissions can accumulate over time, the enclave model ensures that access is tied to a specific data contract. When a user or application no longer needs access to a particular dataset, the data contract can be revoked. This means that access is always temporary and precisely scoped, preventing privilege creep from happening in the first place.
    \item \textbf{Simplifies Auditing}: Because all data access is brokered through a pre-defined data contract, the auditing process becomes significantly more straightforward. Security teams no longer have to audit a chaotic web of individual user permissions. Instead, they can focus on auditing the integrity of the data contracts and the policy engine that governs them. This provides a clear, auditable trail that shows exactly who accessed what data, when, and for what purpose.
\end{itemize}

The Snowflake, Microsoft, and FTX breaches \cite{snowflake_incident}, \cite{microsoft_incident}, \cite{ftx_incident} all highlight the need for this approach. In the Snowflake case, the attackers' lateral movement would have been contained if a compromised user account could only access data within a specific enclave, as defined by a data contract. The Microsoft breach, too, would have been limited if the test account's standing privileges did not extend beyond its isolated enclave, preventing the attacker from creating new OAuth applications.

\section{Conclusion: Transitioning to a Least-Privilege Model}

Breach after breach reveals the same truth: standing permissions are a liability not well suited for the dynamic and distributed nature of the modern cloud. Organizations must adopt a Zero Standing Privilege (ZSP) and Just in Time (JIT) model. 

Attacks like the Retool and FTX incidents \cite{retool_incident}, \cite{ftx_incident} highlight the consequences of granting too much power for too long. In the Retool breach, a single compromised admin account with excessive privileges was used to manipulate customer data. At FTX, a SIM-swap attack on an account with perpetual access to hot wallets led to a complete collapse. These cases make it clear that a multi-layered defense is no longer optional.

For companies holding highly valuable information, such as financial, health, or salary data, the stakes are even higher. By applying ZSP \& JIT principles directly to data records, and not just database connections, you can drastically reduce your attack surface and limit the damage a breach can cause.

Adopting a ZSP and JIT model, including at the data level, means:
\begin{itemize}
    \item \textbf{Embrace Just-in-Time (JIT) and ZSP}: Grant temporary, time-bound access to privileged resources only when absolutely necessary.
    \item \textbf{Enforce Strong Authentication}: Move beyond weak, SMS-based 2FA to more secure methods like passkeys or hardware security keys.
    \item \textbf{Continuously Monitor}: Use automated tools to detect anomalous behavior and unauthorized changes, stopping attacks before they can escalate.
    \item \textbf{Secure Third-Party Access}: Regularly review and restrict the permissions of third-party applications, which are often the weakest link in your security chain.
    \item \textbf{Control Data at the Granular Level}: Implement JIT, ZSP, and least-privilege principles to control access to individual data records, rather than entire database connections.
\end{itemize}

The data enclave architecture is the most effective way to implement this granular, least-privilege approach, especially for enterprises where re-architecting everything is not feasible. This model allows you to contain data exposure with secure, isolated environments, ensuring that access is always limited to the specific data needed for an operation. It's the key to bringing Zero Trust from a network concept to the data level, where it's needed most.

\newpage

% BIB %%%%%%%%%%%%%%%%%%%%%%%%%%%%%%%%%%%%%%%%%%%
\bibliographystyle{unsrt}   
%\bibliography{BIB-File}

% APPENDIX %%%%%%%%%%%%%%%%%%%%%%%%%%%%%%%%%%%%%%
%\appendix
%\input{appendix}

\end{document}